\documentclass[12pt,preprint]{aastex}

\def\gsim{\mathrel{\raise.5ex\hbox{$>$}\mkern-14mu
             \lower0.6ex\hbox{$\sim$}}}

\def\lsim{\mathrel{\raise.3ex\hbox{$<$}\mkern-14mu
             \lower0.6ex\hbox{$\sim$}}}

\slugcomment{Submitted to the Astrophysical Journal.}

\shorttitle{The matter  content of quasar jets}
\shortauthors{Georganopoulos, Kazanas, Perlman, \& Stecker}

\begin{document}

\title{ Bulk Comptonization of the Cosmic Microwave Background by Extragalactic Jets as a Probe of their Matter  Content}

\author{Markos Georganopoulos\altaffilmark{1,2}
, Demosthenes Kazanas\altaffilmark{2}, 
Eric Perlman \altaffilmark{1}, 
Floyd W. Stecker\altaffilmark{2}}

\altaffiltext{1}{Department of Physics, Joint Center for Astrophysics, University of Maryland-Baltimore County, 1000 Hilltop Circle, Baltimore, MD 21250, USA}
\altaffiltext{2}{Laboratory for High Energy Astrophysics, 
NASA Goddard Space Flight Center, 
Code 661, Greenbelt, MD 20771, USA.}

\begin{abstract}

We propose a method for estimating the composition, i.e. the  relative amounts
of leptons and protons, of extragalactic jets which exhibit  
X-ray bright knots in their kpc  scale jets. The method relies on measuring,
or setting upper limits on, the component of the Cosmic Microwave  
Background (CMB) radiation that is bulk-Comptonized by cold electrons 
in the relativistically flowing jet.
These measurements, along with modeling  of the broadband knot emission
that constrain the bulk Lorentz factor $\Gamma$  of the jets, can yield
estimates of the jet power carried  by protons and leptons. We provide 
an explicit calculation of the spectrum of the bulk-Comptonized
(BC)  CMB component and  apply these results to  
PKS 0637--752 and 3C 273, two superluminal quasars with 
 {\sl Chandra} -- detected large scale jets. 
What makes these sources particularly suited for such a procedure is  the 
absence of  significant  non-thermal jet emission in the {\sl `bridge'},
 the region between the core and the first bright jet knot, 
which guarantees that most of the electrons are cold there, 
leaving the BC scattered CMB radiation as the only  significant source 
of photons in this region. 
At $\lambda=3.6-8.0 \; \mu m$, 
the most likely band to observe the BC scattered CMB emission,
 the {\sl Spitzer} 
angular resolution ($\sim 1''-3''$) is considerably smaller than 
the `bridges' of  these jets ($\sim 10''$),  
making it possible to both measure and resolve this emission.

\end{abstract}


\section{Introduction \label{intro}}

One of the primary  issues that impedes our understanding of extragalactic 
jets is that of their composition. 
The  power and momentum transferred by the jets from their point of origin, 
near the central black hole, to the radio 
lobes at Mpc scales is transported by an  electrically neutral, 
yet unspecified 
combination of leptons (both electrons and positrons or $e^\pm$), protons, 
and Poynting flux carried by the  magnetic field of the flow. For a given  
radiative output, the jet power  depends on the composition of the outflowing
matter, with leptonic ($e^\pm$) jets demanding less overall  power and mass  
and being easier to accelerate to relativistic bulk flow velocities
than hadronic ($e-p$) ones. Uncertainty in the matter 
content results in  uncertainty in the jet power, which  bears on our 
understanding of the jet dynamics and the radio lobe energetics, their 
interaction and influence on the host galaxy and possibly the embedding cluster 
core (e.g. Omma \&  Binney 2004; McNamara et al. 2005). The  composition is also  related to  
the fundamental problem of jet formation: the plasma in jets powered by an 
accretion disk (Blandford \& Payne 1982; K\"onigl 1989) would be expected to 
be baryon loaded, while jets powered by  the rotational energy
of a black hole are more likely to result in $e^\pm$ jets (Blandford \& 
Znajek 1977).

A number of attempts have been made over the years toward measuring, or at
the least constraining, the matter content of jets, in particular the 
fraction of kinetic energy stored in protons and low energy or cold leptons, 
whose low radiative efficiencies fail to provide direct evidence of their
presence. To this end, a variety of arguments 
have been presented, based on the synchrotron - self Compton (SSC) 
formalism (e.g. Celotti \& Fabian 1993; Reynolds et al. 1996; Hirotani 2005), 
on the  circular polarization of the radio emission measured in 3C 279 and 
other sources (e.g. Wardle et al. 1998), 
and on the pressure balance of the radio 
lobes and the X-ray emitting confining plasma (e.g.,  Gizani \& Leahy 2004).  These largely indirect arguments have produced rather contradictory results, 
with different works supporting leptonic  and hadronic  
contributions of varying proportions  to the jet content.

A direct estimate of the cold lepton content of  blazar jets  was proposed  
by Begelman \& Sikora (1987), Sikora \& Madejski (2000), and Moderski
et al. (2004). The gist of their argument is the following: The observed 
non-thermal blazar emission is thought to be produced at distances $\sim 10^{17}
-10^{18}$ cm  from the central engine (e.g. Sikora 1997); the jet leptons 
providing the blazar emission at these distances need to be transported 
practically cold by a relativistic flow of bulk Lorentz factor $\Gamma \sim 10$
from the black hole vicinity to the blazar emission site; as these cold jet 
leptons  propagate through the quasar broad line region (BLR) they would  
Compton -- scatter the BLR optical-UV photons (of energy $E_{\rm O-UV} 
\sim 10$ eV) to energies $E_{\rm BC} \simeq \Gamma^2 E_{\rm O-UV}
\simeq 1$ keV, to produce a black--body type hump in their X-ray spectra. 
The fact  that such a feature has not been observed in the inverse-Compton
 dominated X-ray spectrum of blazars was used by the above authors to conclude
that the power in these jets is carried mainly by protons,
 although cold leptons  dominate the number of particles in the jet. 
While this idea is well founded and appealing, concrete answers
are hindered by unknowns such as the  distance  at which the jet is formed, 
its sub-pc scale opening angle and the actual photon energy density of 
the BLR, as well as by the presence of a strong X-ray non-thermal continuum 
that apparently could  ``hide" the proposed bulk-Comptonized component.

Arguments  based on the  bulk Compton (BC) emission used by the above authors
can be applied  to any 
astrophysical site involving relativistic flows. One can then obtain more 
concrete conclusions provided that the flow geometry and the target photon 
density are better determined. Such a site is presented by the large scale 
(100's of kpc) jets of superluminal quasars such as PKS 0637--752 (Schwartz 
et al. 2000; Chartas et al. 2001) and  3C 273 (Marshall et al. 2001; 
Sambruna et al. 2001; Jester et al. 2002) detected  by the {\sl Chandra  
X-ray Observatory}.  Such large scale jets are tightly collimated and propagate through
the   very well understood CMB. 
 The effect of the CMB scattering 
by the jet cold electrons is akin to the well known Sunyaev-Zeldovich (SZ) 
effect associated 
with clusters of  galaxies, with the difference being that in this case 
the CMB photons are scattered by electrons moving with a bulk 
relativistic velocity (see below) rather  than by thermal electrons.

These objects and their jets have morphologies which are conducive to 
applying the arguments referred to above. They exhibit radio, optical, 
and X-ray  emission from the 
quasar core and then from well separated knots along the jet at angular 
distances $\sim 8''$ for  PKS 0637--752  and  $\sim 13''$ for 3C 273. 
The region between the core and the first  knot, which we will refer to 
as the {\sl `bridge'} in the rest of this work,  radiates only very weakly  
in radio, optical, and X-ray energies. This is very important because:  
({\it i}) It shows 
that most of the leptons propagating through the `bridge' are `cold'. 
({\it ii}) 
It provides a region free from unwanted contamination by unrelated  broad 
band non-thermal radiation. 
 As in the SZ effect, where the X-ray cluster properties provide values of
the parameters involved in the scattering of 
the CMB photons, the jet properties  (power and kinematics) are provided
by the properties (spectrum and luminosity) of the {\sl Chandra} --
detected knots of these jets.

Schwartz et al. (2000) noted that {\sl Chandra}-detected X-ray emission from 
the jet knots  
at a projected distance of $\sim 70-100$ kpc from the nucleus of  PKS 
0637--752 is part of a spectral component separate from the knot radio-optical
synchrotron emission  and that it is too bright to be explained through  
SSC emission from electrons in energy  equipartition with the jet magnetic
field.  Observational evidence now indicates that this property is common to 
the jets of many other quasars (Sambruna et al. 2002, 2004; 
Siemiginowska et al. 2002, 2003; Marshall et al. 2005; Jorstad \& Marscher
2004; Yuan et al.  2003; Cheung 2004). 
Tavecchio et al. (2000) and Celotti et al. (2001) proposed that the X-ray
emission is due to external Compton (EC) scattering of CMB photons off 
relativistic electrons in the jet, provided that the jet flow is sufficiently 
relativistic ($\Gamma \sim 10$) to  boost  the 
CMB energy density in the flow frame (by $\Gamma^2$) to the level 
needed  to reproduce the observed X-ray flux. 
This was the first suggestion,
based on extended X-ray emission, that powerful jets retain significantly
relativistic velocities at large distances from the core, a very important
feature because it boosts the level of the  anticipated 
BC emission by $\sim \Gamma^2$.

Here we argue that our recently gained  understanding that the 
{\sl Chandra}-detected quasar jets remain relativistic on scales of 
hundreds of kpc, together with  the IR capabilities of {\sl Spitzer}, 
can be used to measure or  substantially constrain the 
matter content of these jets. In \S \ref{section:bc} we 
calculate the BC spectrum of the CMB as a 
function of the cold lepton kinetic power and Lorentz factor $\Gamma$, 
as well as
its polarization. 
In \S \ref{section:xmech} we present the  mechanisms that have been proposed
to explain the X-ray knot emission, and we argue  that EC off the CMB is the 
most probable mechanism. 
In \S \ref{section:power} 
we use the simple analytical arguments of Dermer \& Atoyan 
(2004a; hereafter DA04) to derive 
the flow velocity and estimate  the number of electrons carried by the jet.
We then apply our calculations to the quasars PKS 0637--752 and 3C 273 in 
\S \ref{appl}.
Finally, in \S \ref{section:discussion} we discuss our findings along 
with some caveats  and present our conclusions.

\section{Bulk Compton Spectrum and Polarization \label{section:bc}}

The expected level of  BC  emission depends on the jet power in cold leptons, 
its length, its Lorentz factor $\Gamma$ and its angle to the observer's line 
of sight. We now proceed to calculate the spectrum and luminosity of the 
radiation produced by the propagation of a collimated cold electron beam 
(mean electron Lorentz factor  $\langle \gamma \rangle \approx 1$) of  
Lorentz factor $\Gamma$ and power $L_{e}$ through an isotropic photon field.
We start with a  simple estimate of the peak energy  and peak luminosity
as a function of observing angle assuming a monoenergetic photon field
with dimensionless energy $\epsilon_0$ (in units of $m_ec^2$)  and
energy density $U$. Assuming that the jet has a length $l$, the fractional
energy loss of a single  electron after propagating this distance $l$ is
\begin{equation}
{\Delta\Gamma\over\Gamma}={4\over 3}\, {\sigma_{\rm T} \over m_e c^2} 
\,\beta \Gamma U l,
\end{equation} 
where $\beta$ is the beam velocity in units of $c$ and $\sigma_{\rm T}$
is the Thomson cross section. Given that the CMB
 photon energy density $U$ scales with redshift $z$ as $U=U_0 (1+z)^4$, 
where $U_0=4.18 \;10^{-13}$ erg cm$^{-3}$, for a
large scale quasar jet propagating through the CMB, the fractional 
energy losses are
\begin{equation}
{\Delta\Gamma\over\Gamma}\approx 1.4 \times 10^{-6} \Gamma_{10}\;   
l_{100 Kpc}\; (1+z)^4.
\end{equation}
Therefore, the bulk deceleration due  to Compton drag off the CMB can be 
safely ignored   for  $z \lesssim 10$, even for the extreme case of a 
purely leptonic large scale jet.  

The inferred isotropic BC luminosity $L_{BC}$ at a given observing 
angle $\theta$ is given by the fractional electron losses multiplied
by  $L_{e}$ and by the beaming pattern $\delta^3/\Gamma$ of a continuous 
flow (e.g. Sikora et al. 1997), 
\begin{equation}
L_{BC}=L_e \; {\Delta\Gamma\over\Gamma} \; {\delta^3\over \Gamma}=
{4\over 3}\, {\sigma_T \over m_e c^2} \; L_e \beta U l \, \delta^3,
\label{eq:LBC}
\end{equation} 
where $\delta=1/[\Gamma(1-\beta\cos\theta)]$ is the well  known Doppler 
factor. The observed luminosity peaks at 
\begin{equation}
\epsilon_{BC}\approx 2 \epsilon_0 \Gamma \delta.
\label{eq:fBC}
\end{equation}
 If we  assume  $\theta=1/\Gamma$, which
corresponds to $\delta=\Gamma$,  the BC  emission has a power
\begin{equation}
L_{BC}\approx 1.4 \times  10^{-4} \;\; l_{100 Kpc} \; \Gamma_{10}^3 (1+z)^4  L_e,
\end{equation}
and  peaks in the IR regime,  $\nu_{BC}\approx 4 \times 10^{13} \; \Gamma^2_{10}$ 
Hz. For constant Lorentz factor jets,  this emission due to the BC scattering
of the CMB will be evenly distributed between the core and the first knot, 
which for  $\theta=1/\Gamma$ will have a projected length $l/\Gamma$, and 
a luminosity per unit projected length scaling as $\Gamma^4$. Note that: 
({\it i}) the BC surface brightness  is independent of $z$, because its 
cosmological decrease by   
$(1+z)^{-4}$ is exactly compensated by the increase of the CMB energy density
by  $(1+z)^{4}$ (Schwartz 2002). ({\it ii}) The observed peak frequency
is also independent of $z$ because, while the CMB  photon energy and 
therefore the peak emission energy scales as $(1+z)$, the received
photon energy scales as $1/(1+z)$.  ({\it iii}) The independence of the 
surface brightness of the BC component on the redshift $z$ and the fact 
that the angular size of an object is roughly constant for $z \gsim 1$, 
imply that the flux of this feature will also remain constant independent of 
the source redshift (Schwartz 2002), in distinction to the core and remaining
 synchrotron 
jet emission  that decrease significantly with increasing $z$.

\begin{figure}
\epsscale{0.8}
\plotone{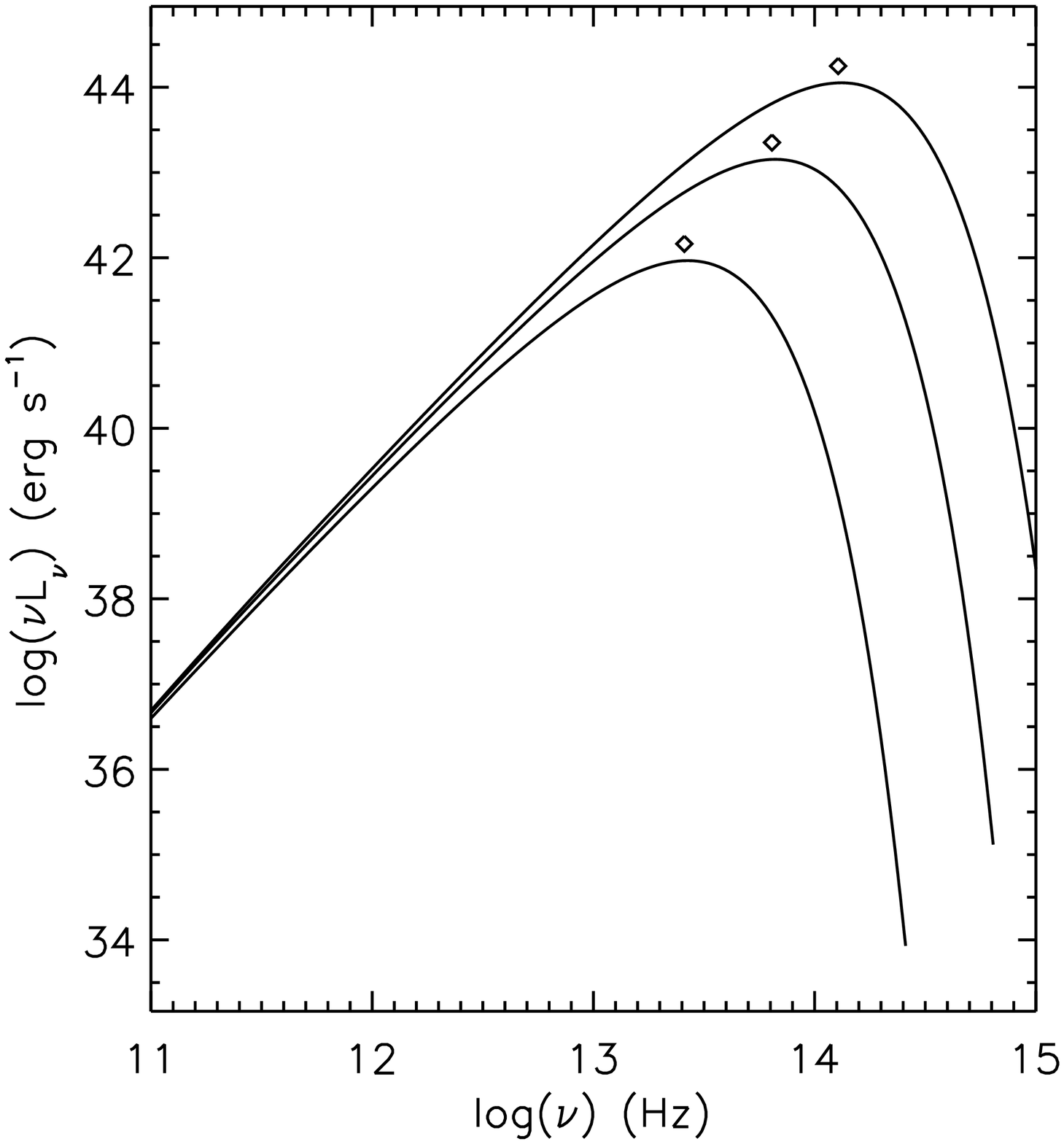}
\caption{The bulk Compton luminosity  of a cold electron beam of power 
$L_e=10^{46}$ erg s$^{-1}$ propagating with a bulk Lorentz factor $\Gamma=10$
for a distance of 100 Kpc, through the CMB at $z=1$. The three curves
correspond to the  luminosity given by Equation (\ref{app.lum}) for observing
angles $\theta=0$ (top curve), $\theta=1/\Gamma$ (middle curve), and $\theta=2/\Gamma$ (bottom curve). The three diamonds correspond to the simple 
power and peak frequency estimates  of Equations  (\ref{eq:LBC}) and 
 (\ref{eq:fBC}).} 
\label{BC}
\end{figure}

{\it The Spectrum.}
To derive the spectrum of the BC  emission off the CMB, we extend 
 the calculation of section \S 7.3 of Rybicki \& Lightman (1979)
from a monoenergetic to a blackbody photon field.  The intensity of the 
blackbody photon field in terms of number of photons is:
\begin{equation}
I(\epsilon)=F_0\; \frac { \epsilon^2 } {\exp(\epsilon /\epsilon_0)-1},
\; F_0 \equiv  \frac{2 m_e^3 c^4}{h^3},\; \epsilon_0 \equiv kT/m_ec^2
\end{equation}
where $T$ is the blackbody temperature. Using the invariance of $I/
\epsilon^2$, we obtain the incident intensity field in the (primed)
frame of the electron beam:
\begin{equation}
I'(\epsilon',\mu'_0)=F_0 \frac{\epsilon'^2}{\exp[\epsilon'\Gamma (1+\beta\mu'_0)/\epsilon_0]-1},
\end{equation}
where $\mu'_0=\cos\theta_0'$ and $\theta_0$ is the angle the incoming
photons form with the direction of the electron motion.
The emission function in the beam frame, assuming isotropic Thomson 
scattering with 
$d\sigma'/d\Omega'_0=\sigma_{\rm T}/4\pi$, is
\begin{equation}
j'(\epsilon_1')=\frac{1}{2}n'\sigma_{\rm T}
\int_{-1}^{+1} I'(\epsilon',\mu'_0) d\mu'_0,
\end{equation}  
where $n'$ is the electron number density in the beam frame.
Performing the integration, we obtain:
\begin{equation}
j'(\epsilon_1')=\frac{n'\sigma_{\rm T}F_0 \epsilon_0}{2\beta\Gamma} 
\epsilon'_1\;
\ln \frac{1-\exp[-\epsilon'_1\Gamma(1+\beta) /\epsilon_0]}
         {1-\exp[-\epsilon'_1 /(\epsilon_0\Gamma(1+\beta))]}.
\end{equation} 
Using the invariance of $j/\epsilon$ we obtain:
\begin{equation}
j(\epsilon_1,\mu)=\frac{\epsilon_1}{\epsilon'_1}j'(\epsilon_1')=
\frac{n'\sigma_{\rm T}F_0 \epsilon_0 }{2\beta\Gamma}\epsilon_1\;
\ln \frac{1-\exp[-\epsilon_1\Gamma^2(1+\beta)(1-\beta\mu) /\epsilon_0]}
         {1-\exp[-\epsilon_1 (1-\beta\mu) /(\epsilon_0(1+\beta))]}.
\end{equation}
  The power of the cold electron beam is $L_e\approx S \beta c \Gamma^2 n'$,
where $S$ is the jet cross section. Using this in the above equation 
and multiplying by the beam volume $V=Sl$, by the final photon energy $\nu=m_ec^2\epsilon_1 /h$, and by $4\pi$, we obtain the BC specific  
luminosity of the source  
\begin{equation}
L_{\nu}=\frac{L_{e}\sigma_{\rm T}l kT}{m_ec^5\beta^2\Gamma^3}\; \nu^2 \;
\ln \frac{1-\exp[-h\nu\Gamma(1+\beta)/(\delta\, kT)]}
         {1-\exp[-h\nu /(\Gamma \delta \, kT(1+\beta))]}, \label{app.lum}
\end{equation}
where we have also used the definition of $\delta$ and the relation
$\epsilon_1/\epsilon_0=h\nu/kT$.
As can be seen in Figure \ref{BC} the simple estimates of the
 luminosity and peak frequency  agree relatively
well with the  results of the spectral calculation, and can be safely used
for simple estimates.

\begin{figure}
\epsscale{0.8}
\plotone{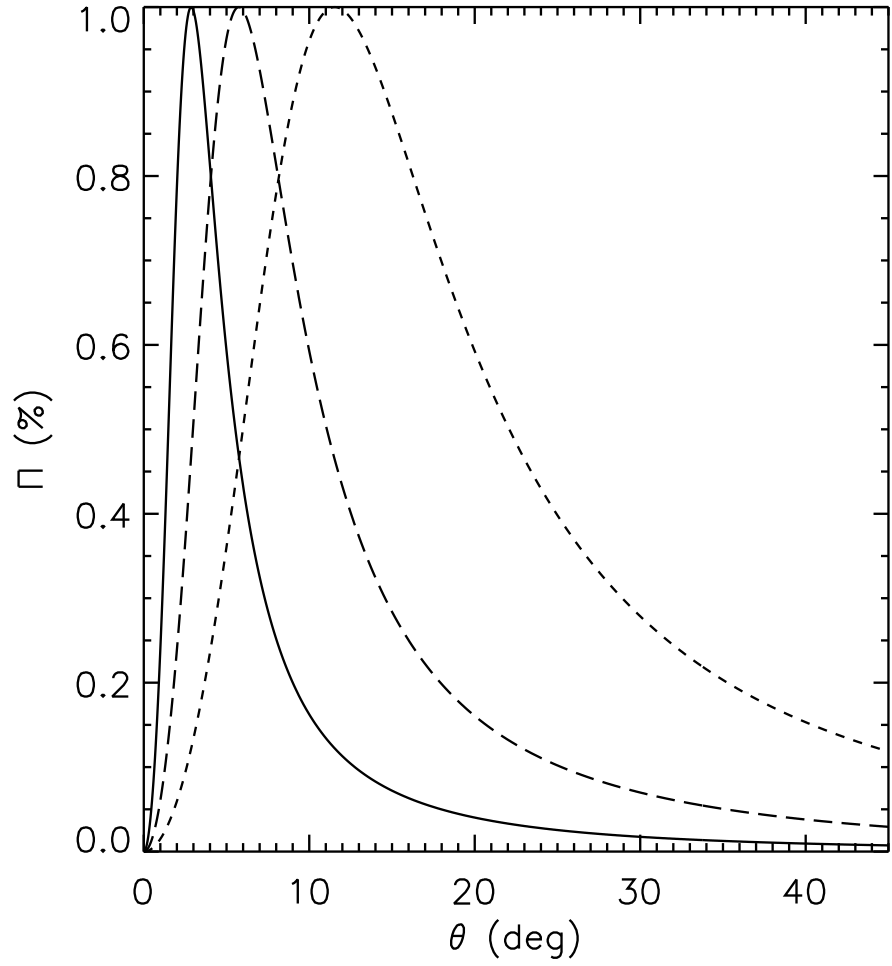}
\caption{The bulk Compton emission polarization from  a cold electron beam
as a function of observing angle for $\Gamma=20$ (solid line), $\Gamma=10$ 
(long dashed line), and $\Gamma=5$ (short dashed line).  }
\label{polosi}
\end{figure}

{\it The Polarization.} The emission we consider, being the result of 
Compton scattering, is expected to be polarized, perhaps very highly so, since
in  the  jet flow rest frame the seed photons are essentially unidirectional.
The polarization can be  easily calculated in the approximation of a 
perfectly collimated cold electron  beam with $\Gamma \gg 1$ (for a general 
treatment  of BC emission polarization
see Begelman \& Sikora 1987). The degree of polarization $\Pi$ is a Lorentz 
invariant and at the beam frame can be written as $\Pi=(1-\mu'^2)/(1+\mu'^2)$. 
Using the light aberration relation $\mu'=(\mu-\beta)\Gamma\delta$, one can 
obtain immediately the degree of polarization $\Pi$ in the observer's frame:
\begin{equation}
\Pi=\frac{1-(\mu-\beta)^2\Gamma^2\delta^2}{1+(\mu-\beta)^2\Gamma^2\delta^2}.
\end{equation}
As can be seen in Figure \ref{polosi}, the degree of polarization is a 
very sensitive function of orientation, reaching $100 \% $ at $\theta=1/
\Gamma$ and dropping  rapidly to $0 \%$ at  $\theta=0$. Using the above 
expression along with the relation $\epsilon_{BC}=2\Gamma\delta\epsilon_0$ 
between the initial $\epsilon_0$ and final $\epsilon_{BC}$ photon energies, one
can obtain the values of both $\beta$ (or equivalently $\Gamma$) 
and $\theta$ in terms of the observables $\Pi$ and $\epsilon_{BC}$. 
Setting $A^2=(1-\Pi)/(1+\Pi)$ and $B=2\epsilon_0/\epsilon_{BC}$, we obtain
\begin{equation}
\beta=\frac{-AB\pm\sqrt{A^2B^2+4(1-B)}}{2}, \;\; \mu=\cos\theta=AB+\beta,
\end{equation} 
where the positive sign corresponds to an approaching beam. Such polarization
measurements of the BC emission can in principle be performed in the near IR
to provide an independent measurement of both the Lorentz factor $\Gamma$ 
and the angle $\theta$ of the jet to the observer's line of sight that can
break the degeneracy normally present in, e.g., VLBI measurements.

\section{The X-ray emission mechanism\label{section:xmech}}

The  BC emission of the CMB in the `bridge' region 
depends on  the power carried by  cold leptons, the flow Lorentz  
factor $\Gamma$ and the angle of the jet to the observer's line of sight 
$\theta$, as shown by Equation (\ref{eq:LBC}). Estimates or constraints 
on these quantities are provided by the non-thermal emission at the  
knot marking the end  of the `bridge'.  The  
{\sl Chandra} detections are critical in  providing such estimates. 
This  brings on the issue of the X-ray emission process in the knots 
of the {\sl Chandra-}detected  quasar jets, as different mechanisms produce 
different  constraints for the quantities on which  the level of BC 
emission depends. While EC of the CMB (Tavecchio et al. 2000; Celotti et
al. 2001) appears the most promising process, we feel that a critical 
examination of the alternatives is at this point necessary.

Dermer \& Atoyan (2002) suggested that the observed X-rays are due to 
synchrotron emission  from electrons cooling by EC off the CMB in the 
Klein--Nishina (KN) regime. Due to the reduced KN-losses of the highest
energy electrons on the thermal CMB spectrum, their distribution function 
develops a ``hump" at these energies, which manifests itself in 
the synchrotron emission of these electrons as an increase in the spectral
luminosity of this component between UV and X-ray energies. The continuity
of the electron distribution function, then, implies also the continuity
of the spectrum between optical, UV, and X-rays. 
As a result,  the extrapolation of the observed X-ray spectrum to lower 
frequencies, namely UV and optical,  must  {\sl always} lie  below the 
observed UV and optical fluxes. 
This is actually contrary
to observation, as indicated by the optical detections  or upper limits 
at several such knots (e.g. PKS 0637--752, Chartas et al. 2000; Knot A of 
1354+195, Sambruna et al. 2004; Knot B of 1150-089, Sambruna et al. 2002;
Knot C4 of 0827+243, Jorstad \& Marscher 2004).  This interpretation
could be valid, however, for  the X-ray 
emission of  knot A of 3C 273, as the near-IR - optical -UV spectrum 
indeed turns  upward, pointing   toward the X-ray point (Jester et al. 2002). 

Alternatively, it has been suggested (e.g. Schwartz 2000) that the 
X-rays are due to synchrotron radiation by a second, very energetic 
electron population with a low energy cut-off at sufficiently high 
energy that its synchrotron emission results in a low energy cut-off
at UV energies to comply with observation. However, even if an unknown
mechanism can produce the injection of an election population with 
the above properties,  
they would cool in less than a knot-crossing time to energies below
those corresponding  to optical synchrotron emission.   This fact has two 
unfavorable implications: ($i$) Given the  observed X-ray spectral indices
($\alpha_x = (\gamma - 1)/2 \sim 0.5 - 0.8 $;  e.g. Sambruna et al. 2004), 
the {\sl injected} electron distribution must have an index $p$ 
flatter by one unit than that observed , i.e., $p = \gamma +1 =2\alpha_x
\sim 1-1.6 $ (see also Aharonian 2002), significantly flatter than 
the asymptotic values  predicted by particle 
acceleration theories  ($ p \simeq 2 - 2.3$, e.g. Kirk et al. 2000). 
($ii$) These high energy  electrons will cool below their low energy 
cut-off to produce, in this energy range, an electron distribution $N_e(
\gamma)\propto \gamma^{-2}$; the synchrotron emission of these electrons
would then lead to a $\nu^{-1/2}$ spectrum that in 
many cases, such as PKS 0637--752, overproduces the observed optical fluxes.
These problems  can be overcome if instead of a simple injection
one invokes the continuous acceleration of the radiating electrons in 
multiple shocks  or spatially distributed stochastic 
acceleration (Stawarz et al 2004). In both cases such models can produce 
a pile-up  of high energy electrons at the upper end of the electron 
distribution which could lead to X-ray synchrotron consistent with
observations. These models, like that of Dermer \& Atoyan (2002), can 
only model successfully emission by sources in which the extrapolation 
of the X-ray spectrum to optical frequencies lies below the observed 
optical flux, such as the knot A of 3C 273.

The EC interpretation of the knot X-ray emission is not entirely without 
problems either. Multiwavelength observations have shown that  in 
many cases the emission profiles at the knot regions are largely achromatic. 
This behavior is unexpected (Tavecchio, Ghisellini, \& Celotti 2003; 
Stawarz et al. 2004), because the cooling length of the EC X-ray
emitting electrons  ($\gamma \sim$ few hundreds) is longer than that 
of the radio  emitting ones ($\gamma \sim$ few thousands) and comparable 
to or longer than the size of the entire jet; this would lead one to 
expect longer jet emission in X-rays than that in the radio. 
%

However, in most {\sl Chandra-}detected quasar jets, the radio-to-X-ray 
logarithmic slope  $\alpha_{rx}$ increases downstream along the jet.  
Indeed, some (e.g. 3C 273 in Sambruna et al. 2001
and  Marshall et al. 2001; PKS 1136--135  and 1354+195  
in Sambruna et al. 2002; PKS 1127--145 in Siemiginowska et al. 2002; 
0827+243 in Jorstad \& Marscher 2004) show anti-correlated X-ray and
 radio maps, with the  X-ray emission peaking closer to the core, 
gradually decreasing  outward, while the radio emission increases outward
 to peak practically  at the jet terminus. 
This problem is alleviated, however, if the  large-scale jet gradually 
decelerates (Georganopoulos \& Kazanas 2004) downstream from the first 
knot. Then,  the X-ray brightness  decreases along the jet because the 
CMB photon energy density in the flow frame decreases.  At the same time, 
the deceleration leads to an increase of the  magnetic field in the 
flow frame, which enhances the radio emission with distance. As a result 
the radio emission is shifted downstream of the X-rays and the  radio 
to X-ray spectral logarithmic slope $\alpha_{rx}$ increases along the jet, 
in agreement with observations. The notion of relativistic and decelerating 
flows in the large scale quasar jets is in agreement with the recent 
suggestion (Georganopoulos \& Kazanas 2003) that the flow in the terminal 
hot spots of powerful  jets must also be mildly relativistic 
($\Gamma\sim 2-3$) and decelerating to sub-relativistic velocities.

In conclusion, we consider the EC model (Tavecchio et al. 2000;
 Celotti et al. 2001), with the modification of the bulk 
flow deceleration  proposed by Georganopoulos \& Kazanas (2004)
 to be the most favorable process accounting for the observed knot
emission in this class of sources. 
However, the synchrotron 
interpretation for the X-ray emission is still viable, at least for
the sources consistent with a continuous underlying electron
distribution, manifest by the continuity between the optical to 
X-ray spectra, as discussed above. This possibility will be considered
in the determination of the jet properties of  3C 273.

\section{The Jet Power \label{section:power}}

Having discussed the mechanism responsible for the X-ray 
emission of the jet knots, we now turn to the determination of their
dynamical parameters, subject to the constraint that they account
for the observed X-ray emission of the knots. These estimates are
based principally on the energetics of the emission rather than the 
details of the spectra, as the latter can be reproduced by appropriate
choice of additional parameters pertaining to the particle distributions.
However, even at this  level, 
the number of parameters exceeds that of the observables. One 
therefore resorts to minimum energy arguments in order to further 
constrain the available values of the magnetic field $B$, the Doppler
factor $\delta$ and the energy flux in or the power of  the jet flows. 

A comprehensive set of  constraints for the jet power and beaming based 
on multiwavelength observations of knots in the extended jet has been recently
presented by DA04 (see also Ghisellini \& Celotti 2001). 
These authors produced analytic relations for the 
jet Doppler factor $\delta_{min}$ that minimizes the jet power and are 
consistent with all constraints imposed by the knot multiwavelength 
emission. In this respect one should note that, because the quantity that is 
minimized is the total knot power and {\sl not} the power in relativistic
electrons and magnetic field only, the value of $\delta_{min}$ depends 
on the matter content of the jet.

\subsection{Minimum Jet Power for Knot X-ray Emission Due to EC}

The work of DA04 models the knots as homogeneous sources moving with a 
Lorentz factor $\Gamma$ at an angle $\theta$ to the line of sight.
The knot matter content  is described through  the ratio $k_{pe}$ 
of the power carried by  protons to the power carried by leptons in 
the knot. For a pure $e^{\pm}$ composition, $k_{pe}=0$, while for an 
$e-p$ jet this parameter reaches its maximum value of 
\begin{equation}
k_{pe}={m_p(p-2)\over m_e \gamma_{min} (p-1)}, 
\label{kpe}
\end{equation}
where $\gamma_{min}$ and $p~ (>2)$ are the minimum Lorentz factor
and energy index of the power law electron energy distribution (EED),
and the protons are considered to be cold in the knot comoving frame
(note that DA04  use   $k_{pe}=m_p/(m_e\gamma_{min})$, valid only for $p\gg1$). 
Assuming that the X-rays are due to EC scattering off the CMB and that 
$\delta=\Gamma$, DA04  calculate  
(their Eq. (12))  the Doppler factor $\delta_{min}$ 
that minimizes the power  that  has to be supplied to the knot by the jet. 
Their expression can be written as
\begin{equation}
\delta_{min}=f_1 \left({1+k_{pe} \over \gamma_{min}^{p-2}}\right)^{1/(5+p)},
\label{delta} 
\end{equation}
where $f_1$  depends  on the source redshift, the radio and
X-ray  fluxes and spectral index (the radio and X-ray index are assumed to 
be the same, as would be the case if they are due to a 
single power-law electron distribution)
and the linear size of the knot ($f_1$, along with the subsequently used $f_2,\; f_3,\; f_4$
are reproduced in Appendix \ref{fs}). 
In particular, as can be seen from Eq. (\ref{kpe}) and  (\ref{delta}), 
the ratio of hadronic to leptonic Doppler 
factors depends only  on $\gamma_{min}$  and $p$:
\begin{equation}
{\delta_{min, e-p}\over \delta_{min,e^{\pm}}}=(1+k_{pe})^{1/(5+p)}\approx
\left[{m_p(p-2)\over m_e \gamma_{min} (p-1)}\right]^{1/(5+p)}, \;\mbox{for} \;
\; k_{pe} \gg 1
\end{equation} 
The lowest  possible value for  $\gamma_{min}$ is constrained by the 
requirement that the EC emission does not extend to frequencies as low
as optical. 
Similarly the maximum possible value for  $\gamma_{min}$ is constrained by the 
requirement to be sufficiently small that its Comptonization of the CMB 
leads to the observed X-ray emission.
In Figure \ref{deltafig} we plot $\delta_{min}$ as a function of $\gamma_{min}$
for the extreme cases of an $e^{\pm}$ (solid curve) and an $e-p$ 
composition  (dashed curve) for the knot WK7.8 of PKS 0637--752. 
The observational parameters needed for the calculation are
taken by Chartas et al. (2000) and Schwartz et al. (2000) and can be found in 
Appendix \ref{fs}.
For those values of $\gamma_{min}$ for which  both solutions are permitted, 
the hadronic knot is  characterized  by  substantially higher $\delta_{min}$.

We now turn to  the minimum knot power $L_{min}$ that corresponds to 
$\delta_{min}$, which, following DA04,  is written as:
\begin{eqnarray}
&L_{min}=L_{part}+L_{B}, \nonumber \\
&L_{part}=f_2(1+k_{pe}) \; \gamma_{min}^{2-p}\;\,\delta_{min}^{-(1+p)} \nonumber\\
&L_B=f_3\;\delta_{min}^4,
\label{eq:power}
\end{eqnarray}
where $L_{part}$ is the power in particles and $L_B$ is magnetic field power,
and $f_2,\; f_3$ are functions  of observables used in DA04 (see  Appendix 
\ref{fs}). In Figure \ref{power}, we plot the total minimum knot power as 
a function of $\gamma_{min}$ for an $e^{\pm}$ (thin solid curve) and an 
$e-p$ (thick solid curve), for the case of knot WK7.8 of PKS 0637--752. 
As can be seen, the minimum power for a hadronic knot flow (equal numbers
of $p$'s and $e$'s) is always larger 
than that for a leptonic one. Interestingly, the leptonic power (dashed line)
needed to produce the observed emission in minimum total power 
conditions is larger for a leptonic jet. This is mostly due to the lower
Doppler factor $\delta$ of leptonic relative to hadronic knot flows and 
the strong dependence of the knot emission on $\delta$.  
In conclusion, {\sl for a given choice of $\gamma_{min}$, the Doppler factor 
and the jet power are only a function of the matter content, and can be 
used to calculate the BC emission using the formalism of  
\S \ref{section:bc}}. 

\subsection{ Minimum Jet Power for  Knot X-ray Emission 
Due to Synchrotron \label{synch}}

If we assume a synchrotron interpretation for
the broadband knot spectrum, then, as DA04 point out, 
the power needed to produce  the second high energy component is 
only a small fraction of the power 
needed to produce the low energy synchrotron emission. This is mostly because
the electron radiative efficiency is much smaller at radio energies.
Therefore, we focus on the power needed to produce the radio emission.
Here,   for a given observed synchrotron radio spectrum,
the Doppler  factor at minimum  power conditions cannot be uniquely defined.
Instead, it is    the product $B\delta$ that can be derived.
Following DA04, their Eq. (8) can be written as 
\begin{equation}
\delta\epsilon_B=f_4\left( { \gamma_{min}^{2-p}\over \tilde{k}_{eq}}\right)^
{2/(5+p)}, \label{f4}
\end{equation}
where $\epsilon_B$ is the magnetic field in units of the critical magnetic 
field $B_{cr}=m_e^2c^3/e\hbar $,  $f_4$ is a function of observables
(Appendix \ref{fs}), and $\tilde{k}_{eq}$ is the ratio of leptonic
to magnetic field energy density in the knot. For $\tilde{k}_{eq}=1$, 
$y_{eq}=(\delta\epsilon_B)_{eq}= f_4   \gamma_{min}^{2(2-p)/(5+p)}$ and 
 $\delta\epsilon_B=y_{eq}/\tilde{k}_{eq}^{2/(5+p)}$.
The jet power flowing through the knot  is  
$L=\pi r_b^2 c \delta^2(u_{part}+u_B)$, where $u_B$ is the magnetic field
energy density, $u_{part}=(1+k_{pe})\tilde{k}_{eq}u_B$ is the total particle
energy density, and we have made the usual assumption $\delta=\Gamma$. 
The jet power can then be written as
\begin{equation}
L=\pi r_b^2 c u_{B_{cr}} [(1+k_{pe})y_{eq}^{(5+p)/2}y^{-(1+p)/2}+y^2],
\label{eq:lum}
\end{equation}
where $y=\delta\epsilon_B$  and  it is minimized for 
\begin{equation}
y_{min}=(\delta\epsilon_B)_{min}=f_4   \gamma_{min}^{-2(p-2)/(5+p)}\left[(1+k_{pe})(p+1)\over 4 \right]^{2/(5+p)}.\label{eq:ymin}
\end{equation}
Setting $y=y_{min}$ in Equation (\ref{eq:lum}) we can calculate the jet power
as a function of $\gamma_{min}$ and the jet matter content expressed through $k_{pe}$.

\section{Applications\label{appl}}

We focus our attention on the  {\sl Chandra}-detected  superluminal quasars  
 PKS 0637--752 and 3C 273, whose  superluminal nature guarantees that their 
jets must be at  relatively small viewing angles  $\theta\sim 1/\Gamma$, where
 $\Gamma\approx \beta_{app}$ is the minimum
bulk Lorentz factor that corresponds to the  detected superluminal velocity 
$ \beta_{app}$ (see e.g. Urry \& Padovani 1995). This suggests  actual 
`bridge' lengths $\sim \Gamma$ times longer than their projections on 
the plane of the sky, resulting in  separations of hundreds of kpc between 
the core and the first knot.

Using the analytic expressions of \S  \ref{section:power} we first estimate  
the power $L_{lept}$ in leptons and Doppler factor  $\delta$ required 
to reproduce the emission from the knot at which the radiatively inefficient 
`bridge' terminates, under minimum power conditions.
We then proceed to calculate the BC flux  using the formalism
of \S \ref{section:bc} under two different, general assumptions. 

 {\sl Case A:} 
The lepton
power $L_{lept}$ required in the knot is provided by the cold leptons in the 
flow ($L_{e}=L_{lept}$). This then requires that only a minority of these leptons
get accelerated at the knot to create the X-ray producing population by tapping
a small fraction of the kinetic energy of the remaining `cold' leptons. 
This represents one of the most optimistic cases for detecting the BC CMB emission
in the `bridge' region as it requires a large number of cold leptons in the 
jet. 

{\sl Case B:} The most conservative case for detecting the anticipated
 BC emission is that in which the jet provides only the number of 
leptons needed to produce the observed X-ray emission at knot; the leptons 
are accelerated there using exclusively the energy of other agents such 
as  the magnetic field and/or the jet hadrons.  In this case  $L_{e}=
L_{lept}(p-2)/(p-1)\gamma_{min}$, and the cold  lepton luminosity, 
and subsequently the BC emission is lower by a factor of $\gamma_{min}
(p-1)/(p-2)$ compared to case A. 

In both cases the jet composition can range from purely leptonic to 
equal number of electrons and protons. In the following we discuss
only these two extreme cases as they bracket all other combinations
of protons and leptons. 
In both cases we assume that the Doppler  factor of the flow in the 
`bridge' is not significantly different from that derived for the knot. 
We defer a discussion of  this assumption for  \S 
\ref{section:discussion}.

\subsection{PKS 0637--752\label{section:0637}}

\begin{figure}
\epsscale{0.8}
\plotone{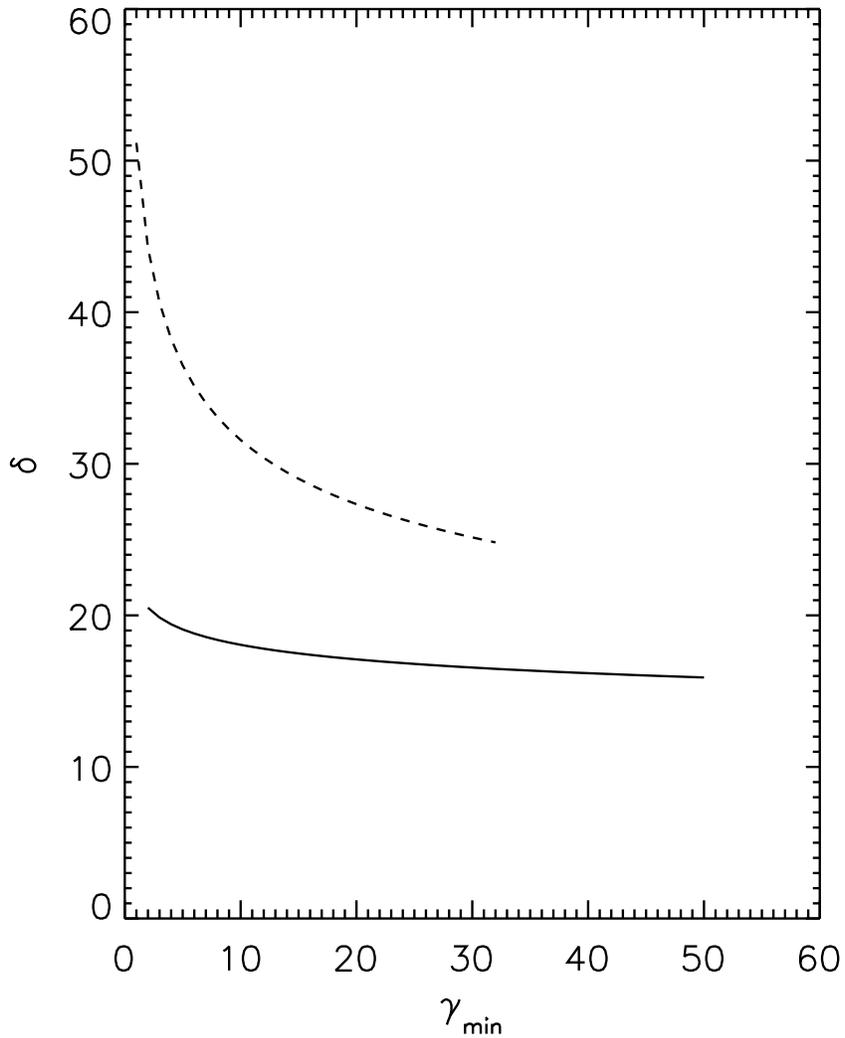}
\caption{The Doppler factor $\delta_{min}$ that minimizes the power required by
 the knot  Wk7.8 of PKS 0637--752 as a function of the permitted values of
the Lorentz factor $\gamma_{min}$ of the EED low energy cutoff.
The solid curve corresponds to  an $e^{\pm}$ composition, while the dashed
curve to an $e-p$ composition. The observational data used in calculating
 $f_1$ in Equation (\ref{delta}) are those used by DA04 and are taken by 
Schwartz  et al. (2000) and Chartas et al. (2001).  Note that the range 
of permitted   $\gamma_{min}$    is  a function of the knot matter content.}
\label{deltafig}
\end{figure}

 In the case of PKS 0637--752, the optical flux from knot WK7.8, the knot
at which the radiatively inefficient `bridge' terminates, is clearly below
the extrapolation of the X-ray spectrum at lower frequencies, and, 
as we argued in \S \ref{section:xmech}, this favors the EC 
interpretation for the X-rays. 
For  knot WK7.8 we adopt $\gamma_{min}=20$, which, using Equation 
(\ref{delta}) and the observational data in Appendix \ref{fs},
  corresponds to  minimum
power Doppler factors $\delta_{min}=17.4$ for a leptonic composition 
and   $\delta_{min}=27.8$ for a hadronic composition, 
as can be seen in Figure \ref{deltafig}. 
It also   corresponds to a jet 
minimum power $L_{min}=9.7 \times  10^{45} $ erg s$^{-1}$ for the leptonic
jet and  $L_{min}=6.3 \times  10^{46} $ erg s$^{-1}$ for the hadronic jet.
The  corresponding lepton power is  $L_{lept}=3.7 \times 10^{45} $ 
erg s$^{-1}$ for the leptonic
jet and  $L_{lept}=6.8 \times  10^{44} $ erg s$^{-1}$ for the hadronic one.
As can be seen in Figures \ref{deltafig} and \ref{power}, 
these numbers are only weakly affected by our choice of $\gamma_{min}$,
as long as $\gamma_{min}\gtrsim 10$.
To estimate the actual length of the `bridge',
we make the usual for superluminal sources assumption that the source is observed at an angle
  $\theta=1/\Gamma$. 
At the redshift of the source ($z=0.651$), 
 $1''$ corresponds to $6.9$ Kpc assuming standard cosmology
(Spergel et al. 2003) and  the deprojected `bridge' length
 is $l \approx $ 930 Kpc for a leptonic jet and $l \approx  1.5$
 Mpc for a hadronic jet. We are discussing the implications of these length
estimates in \S \ref{section:discussion}.

\begin{figure}
\epsscale{0.8}
\plotone{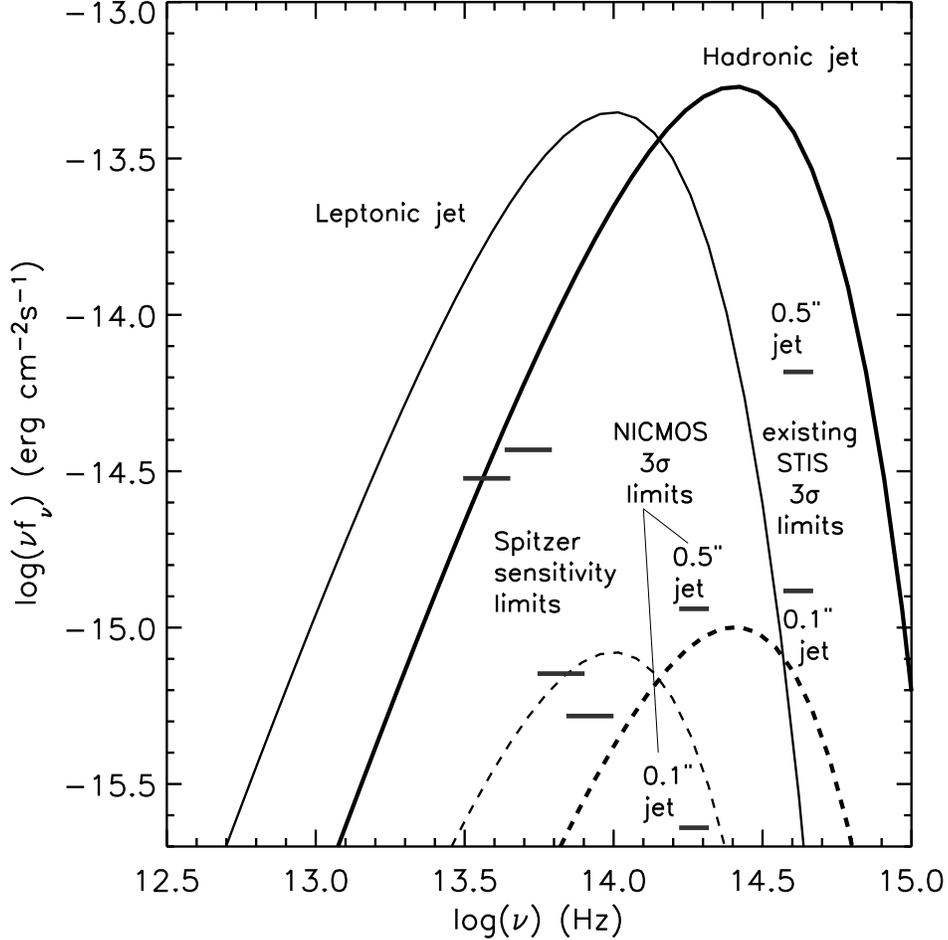}
\caption{The BC emission for a leptonic (hadronic) jet composition of
 PKS 0637--752  is plotted with a solid thin (thick) line for case A,  
in which the lepton power $L_{lept}$ 
required in the knot is provided by the cold leptons in the beam. 
The dashed lines correspond to case B, in  which the jet provides
 simply the number of  leptons needed in the knot, with the thin (thick) line 
representing a leptonic (hadronic) jet composition. The {\sl Spitzer} 
sensitivity limits, existing $3\sigma$ STIS {\sl HST} limits (see text),
and expected $3\sigma$ NICMOS {\sl HST} $\lambda=1.6 \mu m$ limits for
 a 3-orbit  exposure,  assuming a $0.1''$ or a $0.5''$  jet radius are also 
shown.}
\label{0637}
\end{figure}

To calculate the BC flux we  proceed using the formalism 
of \S \ref{section:bc} for an $e-p$ and an $e^{\pm}$ composition
for both cases A and B for the cold lepton power described above.
As can be seen in Figure \ref{0637}, in case A 
the emission for  a leptonic jet peaks at mid IR energies,
while that for a hadronic jet peaks at near IR - optical energies. 
For both compositions  the anticipated mid IR flux is above the {\sl Spitzer} 
sensitivity limits; the hadronic case however violates the {\it HST} 3$\sigma$
detection limits  for both a $0.5''$ and $0.1''$
thin jet.  These limits are derived from 2001  STIS observations at 7219 \AA,
  P.I. Meg Urry.
In the second, most conservative case, the BC emission is still above
the {\sl Spitzer} sensitivity limit for the two shorter wavelength bands.
However, the existing {\it HST} optical limits cannot be used to argue 
against a hadronic jet in this case. 
The angular length of the `bridge'  of PKS 0637--752 is $\sim 8''$. This is 
easily resolved by {\sl HST}. Most importantly, at $\lambda=3.6-8.0 \; \mu m$, 
the most likely band for the BC scattered emission to appear,
the {\sl Spitzer}  angular resolution ($\sim 1''-3''$) is considerably
smaller than the `bridge' size, and we anticipate that  {\sl Spitzer}
will resolve the BC emission along the `bridge'.   Additional constraints 
on the BC SED can be imposed by NICMOS {\ sl HST} observations at 
$\lambda=1.6 \mu m$ (see  Figure \ref{0637}).

\subsection{3C 273 \label{section:273}}

The X-ray emission from  knot A of 3C 273 has been interpreted
as both synchrotron (Marshall et al. 2001) and EC off the CMB
 (Sambruna et al. 2001). The discrepancy can be, at least partially,
resolved by studying the  spectrum of  the near IR - optical - UV emission.
{\sl HST} and {\it VLA} observations by Jester et al. (2002) showed that
the near IR - optical - UV spectrum of knot A is flatter than the radio - 
near IR spectrum, indicating the presence of a high energy component, which, 
they suggest, can be interpreted either as EC off the CMB or as a second,
synchrotron component from an independent high energy electron population.
The  constraints on the jet power 
and matter content depend on   the  interpretation we adopt.

{\sl EC off the CMB.} 
We examine first the possibility that the X-rays in knot A 
are due to EC off the CMB. We adopt again $\gamma_{min}=20$, which, 
using Equation (\ref{delta})  and data from Marshall et al. (2001) 
listed in Appendix \ref{fs}, corresponds to a minimum
power Doppler factor $\delta_{min}=16.6$ for a leptonic and 
 $\delta_{min}=26.5$ for a hadronic composition (these values
of $\delta$ are significantly higher those inferred from observations
of superluminal motions ( $\delta\sim 10$; Pearson et al. 1981)  assuming
$\delta=\Gamma$, 
a problem we discuss in \S \ref{section:discussion}). 
This  corresponds to a jet 
minimum power $L_{min}=3.3 \times  10^{45} $ erg s$^{-1}$ for a leptonic
jet and  $L_{min}=2.1 \times  10^{46} $ erg s$^{-1}$ for a hadronic jet.
The  corresponding lepton power is  $L_{lept}=1.6 \times 10^{45} $ erg s$^{-1}$
 for a leptonic
jet and  $L_{lept}=2.9 \times  10^{44} $ erg s$^{-1}$ for a hadronic jet.
As in the case of PKS 0637--752, these numbers are only weakly affected by 
our choice of $\gamma_{min}$, as long as $\gamma_{min}\gtrsim 10$.
The deprojected `bridge' length, derived under the 
assumption   $\theta=1/\Gamma$ (at the redshift of the source, $z=0.158$, 
 $1''$ corresponds to $2.7$ Kpc)  is $l \approx $ 580 Kpc for a leptonic jet 
and $l \approx 930 $ Kpc for a hadronic jet. 

\begin{figure}
\epsscale{0.8}
\plotone{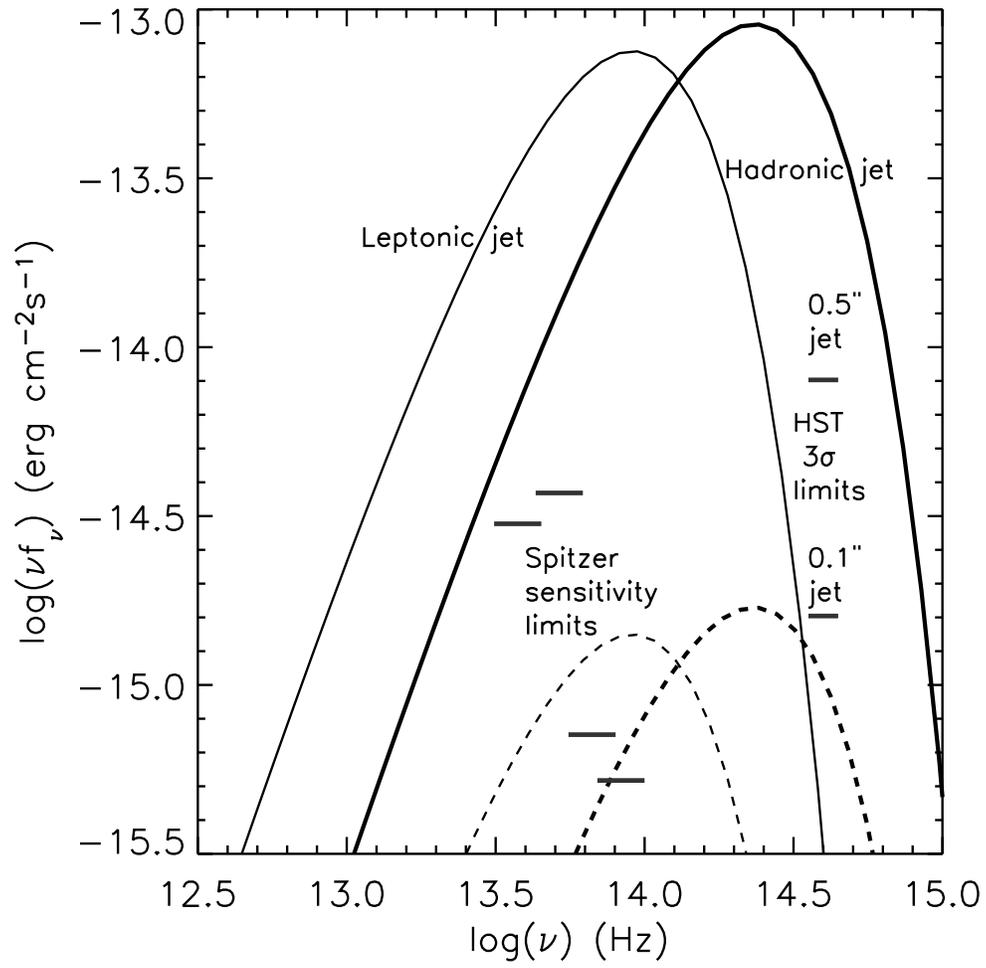}
\caption{Same as in Figure \ref{0637} for  3C 273. ACS {\sl HST} limits 
derived from Martel et al. (2003).}
\label{273}
\end{figure}

Using these values for the Doppler factor and `bridge' length 
we  calculate the BC flux 
 for an $e-p$ and an $e^{\pm}$ composition
for both cases A and B for the cold lepton power.
As can be seen in Figure \ref{273}, the situation is similar with that
in  PKS 0637--752. 
The BC emission of  a  leptonic jet peaks at mid IR energies,
while that of a  hadronic jet at near IR - optical energies. 
In case A, the anticipated mid IR flux is well above the {\sl Spitzer} 
sensitivity limits for both compositions; 
the hadronic composition for case A, however, violates the {\it HST} 3$\sigma$
detection limits  for both a $0.5''$ and $0.1''$
thin jet. 
In the most conservative case B, the BC emission is still above
the {\sl Spitzer} sensitivity limit for the two shorter wavelength bands.
The expected optical emission is lower than the {\sl HST} 3$\sigma$
detection limits for both compositions, offering no additional constraints.
As in PKS 0637--752 , the large angular size of the `bridge' ($\sim 13''$) 
guarantees that the BC emission of the `bridge' can be resolved 
by {\sl Spitzer}.

\begin{figure}
\epsscale{0.8}
\plotone{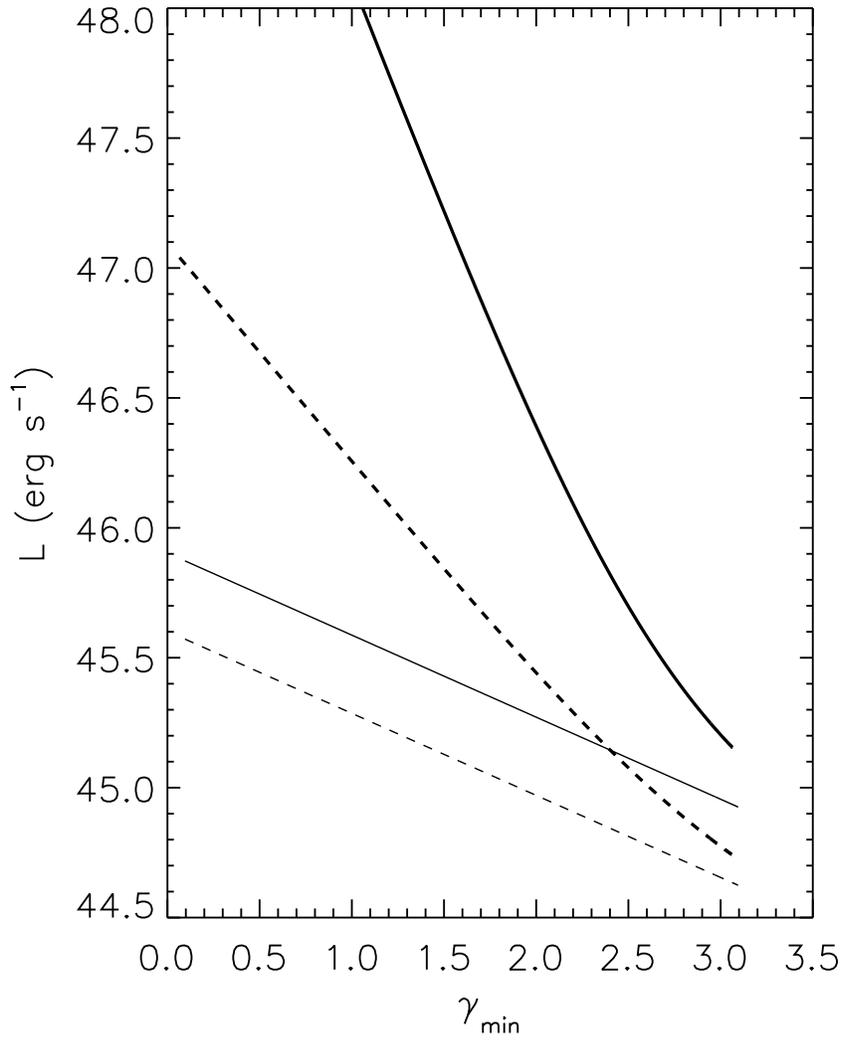}
\caption{The total power required to reproduce the synchrotron emission
of knot A is 3C 273 is shown as a thin (thick) solid line for a leptonic (hadronic) jet composition. The thin (thick) broken line corresponds to the leptonic power for a leptonic (hadronic) composition.}
\label{synch273}
\end{figure}

{\sl Synchrotron.} We turn now to the synchrotron interpretation.
Using Equations (\ref{eq:lum}, \ref{eq:ymin}), we 
calculate the jet minimum power as a function of $\gamma_{min}$. 
An upper limit on $\gamma_{min}$ is set by the requirement that the 
lowest energy  electrons are energetic enough to produce the lowest observed
synchrotron emission
at $\nu_{s,min}=408 $ MHz (Foley \& Davis 1985) associated with knot A: 
$ h \nu_{s,min}/m_ec^2 < y_{min}\gamma_{min}^2(1+z)$. 
As can be seen in  Figure \ref{synch273}, in the case of a leptonic
composition the minimum power required in the leptonic component 
(thin dashed line) 
drops as  $\gamma_{min}$ increases, down to  
$\approx 4\times 10^{44}$ erg s$^{-1}$, a factor of $\approx 4$ below the 
  minimum power in the leptonic component under  the EC interpretation.
In the case of a hadronic composition, the minimum power required in the leptonic component 
(thick dashed line) also decreases with  $\gamma_{min}$; even at its lowest
value, however, it remains more powerful than the  the minimum power
in the leptonic component under  the EC interpretation.
Note that the difference in leptonic power 
between the leptonic and the hadronic compositions decreases with
increasing $\gamma_{min}$ as the energy per electron gradually becomes
comparable to the proton rest mass energy.

Calculating the BC component in the synchrotron case requires a choice
of $\delta$.
Small values of $\delta$ will render the BC emission undetectable, since
$L_{BC}\propto \delta^3$.
Adopting the same values for $\delta$ as those derived in the 
EC off the CMB case, results in  detectable by {\sl Spitzer} BC emission for both leptonic
and hadronic jet compositions in case A.
In case B, the 
BC flux  can drop below the {\sl Spitzer} detectability limits, because the
large permitted values of $\gamma_{min}$ reduce the power of the cold lepton
beam ( $L_{e}=L_{lept}(p-2)/(p-1)\gamma_{min}$).

\section{Discussion and  Conclusions \label{section:discussion}}

In the preceding sections we have formulated and examined the process 
of bulk Comptonization of the CMB photons by the `cold' electrons of the 
relativistic flows of {\sl Chandra-}detected extragalactic jets. 


The physical process of BC scattering of the CMB  is certainly taking place,
as long as there are cold electrons propagating in a jet. One, however, has 
to focus on systems where the BC signature  is expected to be
($i$) strong  and ($ii$) minimally contaminated by  other emissions. 
We argued that these conditions are  favorably  met in  
the {\sl Chandra} - detected superluminal quasars PKS 0637--752 and 3C 273, 
sources in which the jet radiates very weakly  in  radio, optical, and X-ray 
energies for  $\sim 10''$  between the core and the first knot. Based on the
fact that the `bridge' connecting the core to the first knot is only weakly
radiating, we argued that most of the leptons in this radiatively inefficient
 section of the jet are transported practically cold. We then calculated the 
power and Doppler factor of the flow required in the first knot to  produce 
the observed broadband spectrum under minimum energy conditions (DA04),  
adopting EC scattering of the CMB
by   relativistically moving plasma as the X-ray emission mechanism.
For 3C 273 we also examined the possibility that the X-ray emission
is due to synchrotron, a viable alternative for this source.

Using these power and Doppler factor estimates,
we calculated the BC emission for an  $e-p$ and $e^{\pm}$  jet composition,
in each case considering two ways for energizing the electrons in the knot:
in case A we assumed that  the lepton
power needed in the knot is provided by the cold leptons in the 
beam alone, while in the most conservative case B that  
the jet provides simply the number of  leptons needed to produce the 
knot emission (while the required power is provided by another agent,
i.e. protons, magnetic fields).
The resulting BC mid IR emission is above the {\sl Spitzer} 
detectability limits in both cases and for both compositions, 
and actual  {\sl Spitzer} measurements of the `bridge' mid IR emission, 
together with  optical - near IR observations, 
 possibly including  near IR polarimetry will  
 measure or, at worst, substantially constrain the 
matter content of these jets.  As we showed in \S  \ref{section:bc}, 
a measurement of the polarization of the BC component,
 together with an estimate of its peak frequency, can  break the 
degeneracy between the orientation of the jet and its bulk Lorentz factor.

Existing {\sl HST} limits for both 3C 273 and PKS 0637--752 already 
disfavor  case A  $e-p$ models, in agreement with similar  conclusions from
blazar studies  Sikora \& Madejski (2000).
Additional constraints for  pure $e-p$ jets come from the 
large Lorentz factors required. 
Although values of  $\delta_{min}\sim 30$  are still compatible with the 
apparent superluminal motions observed in   some blazars 
(e.g. Jorstad et al. 2002),  the number of such 
highly relativistic sources should not overproduce the  
 parent  (misaligned) source population (e.g. Lister 2003).
 Additionally, as we mentioned in \S  \ref{section:0637}, 
\ref{section:273}, 
the large Doppler factors required for   pure $e-p$ jets, suggest jet lengths
  over $1$ Mpc long, a value barely compatible with 
the largest jets of known radio galaxies 
(e.g. Subrahmanyan, Saripalli, \& Hunstead 1996).

Measuring the BC emission of the 'bridge' can be used to measure
what fraction of the cold electrons propagating in the 'bridge' are picked up
by the particle acceleration mechanism in the knot and are accelerated to
 high energies. This ``injection efficiency'' of particle acceleration
is, so far,   a theoretically not well understood and observationally not
  strongly constrained quantity  (e.g. Gallant 2002). 

A failure to detect the BC emission has definite implications for the matter
content of jets; however, given the limited sensitivity of {\it
Spitzer}, such a nondetection will leave several possibilities open.  For 3C
273, where a synchrotron interpretation of the knot X-ray emission is a
plausible alternative,  a non-detection would still be compatible with 
synchrotron X-ray emission in Case B.
A possible source of the separate high energy electron population required
could be  the decay of a neutron beam (Dermer \& Atoyan 2004b).
If the knot X-ray emission is EC in nature (this
mechanism is plausible for 3C 273 and favored for PKS 0637--752, as we discuss
in \S \ref{appl}),
deviations from equipartition (e.g. Kataoka \& Stawarz 2004) and/or
from the adopted orientation $\theta=1/\Gamma$ can reduce the BC flux below
the {\sl Spitzer} detection limits in case B, in which case we will be able
set an upper limit on the number of cold leptons in the jet as a function
of $\delta$. 
 We note here that the minimum power needed in the EC
model is already  $ \gtrsim 10^{46}$  erg s$^{-1}$  and one cannot deviate 
significantly from that without requiring jet powers
greater than the Eddington luminosity of a $10^9 \; M_{\odot} $   black hole.
If significant deviations from equipartition are needed to explain future
non-detection of the BC emission by {\sl Spitzer}, they will impose severe 
constraints on the EC knot emission model, 
particularly in the case of the higher power hadronic jets.
Even if {\sl Spitzer} does not detect the BC emission, our method can still 
be applied in the near future using the {\sl JWST}; this instrument is
expected to be $\sim 3$ orders of magnitude more sensitive than {\sl Spitzer} 
and therefore probe much fainter levels of BC emission.
Additional constraints on the BC bridge emission can also be provided by  
$\lambda=1.6 \mu m$ NICMOS {\sl HST} observations (see Figure \ref{0637}).

An assumption made in our calculations is that the Doppler factor of the 
jet flow in the `bridge' between the core and the knot is the same as the
Doppler factor of the knot. This can happen if the  flow
does not decelerate substantially at the knot, which can happen 
if the knot is the site of an oblique shock. 
 Note, however, that if the knot is a separate entity propagating 
in the jet, as in one of the  cases  examined by Tavecchio et al. (2003)
 and  Stawarz et al. (2004), the  physical properties of the knot 
will be unrelated to those of the ``bridge" and our method will  not 
applicable for the determination of any of the jet parameters.
In the case of PKS 0637--752,
VLBI observations of superluminal velocities with $v_{app}=17.8\pm 1 \,c $
 in the core of the source (Lovell et al. 2000)
set limits for $\Gamma>17.8$, $\theta<6^{\circ}.4$, in agreement with the
 Doppler factor $\delta=17.4$ derived from minimizing the jet power in an
$e^{\pm}$ jet. Similar values for the Doppler factors in the core and 
the first knot were also derived by Tavecchio et al. (2004) using spectral 
modeling of both the core and the first knot for blazars PKS 1510-089 and 
1641+399 and by Jorstad \& Marscher 2004 for the superluminal
source  0827+243; these authors concluded that there is no bulk flow
 deceleration  between the core and the  first knot in these sources.  
The situation,  however, seems different for  3C 273, where the superluminal
velocities observed in the VLBI core suggest $\Gamma\sim 10$ (e.g. Pearson et
 al. 1981), significantly  lower than  the Doppler  factor $\delta=16.6$  
needed for an $e^{\pm}$ jet in minimum power. This leaves open the 
possibility for a synchrotron interpretation of the X-ray emission, 
which, as we argued in \S \ref{synch}, \ref{section:273}, does not allow 
for a unique determination of $\delta$ and, subsequently, for a firm 
estimate of the BC emission.

We have  also assumed in our calculations the leptons in the radiatively 
inefficient `bridge' between the core and the first knot are cold ($\langle 
\gamma \rangle \approx 1$).  Considering that electrons in the large 
scale jet cannot cool down to $\gamma \sim 1$ in the flow frame once 
accelerated to relativistic energies, the  cold  electrons in the ``bridge'' 
region, in the situation we envisage, are cold  not because they have been
radiatively or adiabatically cooled down, but because they have never been
accelerated. In fact, this is in agreement with our understanding of 
stochastic particle acceleration, according to which, generally,  only a 
small fraction of the available particles is accelerated to high energies.
Clearly there are non-thermal electrons, since the broadband
emission in the `bridge' of both PKS 0637--752 and 3C 273 is weak  but not 
absent. However the majority of the electrons must be at Lorentz factors 
smaller than those required to obtain EC off the CMB at optical energies.
For $\Gamma\approx 10 $ this implies $\gamma \lesssim 4 $.
If the electrons  are not really cold,
but they have a distribution around some small  $\gamma$,  because 
the electron losses scale as $\gamma^2$, the power and the peak frequency 
of the bulk Compton emission would be higher by the same factor.
Our assumption, therefore, represents  a lower limit on the expected  
BC emission.

As was discussed by Schwartz (2002) the knot X-ray emission due to EC
 off  the CMB will remain visible at the same flux level independent of 
redshift. This is also the case for the BC emission from cold leptons 
in relativistic
jets. This suggests an exciting possibility for  jets that have
a very low radiative efficiency past the core (practically sources like 
PKS 0637--752, but without the acceleration events that produce the broadband
non-thermal knot emission): their IR-optical  BC emission will be detectable
independent of redshift, and it will be the only observable signature of 
these otherwise invisible jets.

\appendix
\section{Functions and observables \label{fs}}

We reproduce here the functions $f_1, \, f_2, f_3, f_4$ that result from
 the formalism of DA04.  The function $f_1$ appears in Equation (\ref{delta})
- eq. (12) of DA04 - that relates the minimum power Doppler factor 
$\delta_{min}$ to the knot 
matter content as expressed through $k_{pe}$ and the minimum Lorentz factor
$\gamma_{min}$ of the electron distribution: 
\begin{eqnarray}
& f_1=\left({9 (1+p) d_L^2 m_e c^2(1+z)^{(p-3)/2}      \over    
8 (p-2)c\sigma_{\tau}r_b^3}\right)^{1/(5+p)} 
 \left[ (f_{\epsilon_{EC}}^{EC})^{5+p}u_{B_{cr}}^{3-p} \over 
(f_{\epsilon_s}^s)^4 u_{\star}^{p+5}\right]^{1/(1+p)(5+p)} \times \nonumber \\
& \epsilon_s^{2(3-p)/(p+1)(p+5)} \times 
\left(2\epsilon_{\star}\over \epsilon_{EC}\right)^{(3-p)/2/(p+1)}.
\end{eqnarray}
$f_2$ arises in the calculation of minimum jet power, by
combining eq. (11) of DA04 for the particle energy content with the expression
$L_{part}=\pi r_b^2\Gamma^2c W'_{part}$ and setting $\Gamma=\delta=\delta_{min}$:
\begin{equation}
f_2={9\pi m_ec^2 d_L^2f_{\epsilon_{EC}}^{EC}  \over 2 \sigma_{\tau}u_{\star}r_b(p-2)}
\left( (1+z)\epsilon_{EC}\over 2\epsilon_{\star}\right)^{(p-3)/2}.
\end{equation}
$f_3$ arises in the calculation of minimum jet power, by
combining eq. (10) of DA04 for the  magnetic field with the expression
$L_{B}=\pi r_b^2\Gamma^2c W'_{B}$ and setting $\Gamma=\delta=\delta_{min}$:
\begin{equation}
f_3=\pi r_b^2 c u_{B_{cr}} \left( f_{\epsilon_s}^s u_{\star} \over 
f_{\epsilon_{EC}}^{EC}u_{B_{cr}}\right)^{4/(p+1)}
\left(\epsilon_{EC} \over 2 \epsilon_s \epsilon_{\star}\right)^{2(3-p)/(p+1)}.
\end{equation}
$f_4$ appears in Equation \ref{f4} (eq. (8) of DA04):
\begin{equation}
f_4=\left[ 9 m_e c^2 d_L^2 f_{es}^s \epsilon_s^{(p-3)/2}\over 
2 c \sigma_{\tau} u_{B_{cr}}^2 (p-2) r_b^3\right]^{2/(5+p)}.
\end{equation} 
These  functions depend on  the following quantities:\\
$z$, the source redshift\\
$d_L$, the source luminosity distance\\
$p> 2$, the electron index, related to the radio and X-ray spectral index $\alpha=(p-1)/2$\\
$r_b$, the knot radius\\
$u_{B_{cr}}=B_{cr}^2/8\pi=7.75\; 10^{25}$ erg cm$^{-3}$, where $B_{cr}=m_e^2c^3/e\hbar$ is the
critical magnetic field\\
$u_{\star}=4\,10^{-13} (1+z)^4$ erg s$^{-1}$, the CMB photon energy density at
$z$\\
$\epsilon_{\star}=2.70kT_{CMB}(1+z)/m_ec^2=1.24\,10^{-9}(1+z)$, the dimensionless CMB photon energy at $z$ \\
$\epsilon_s=h\nu_s/m_ec^2$, where $\nu_s$ is the radio (synchrotron) observation frequency\\
$\epsilon_{EC}=h\nu_{EC}/m_ec^2$, where $\nu_{EC}$ is the X-ray (EC) observation frequency\\
$f_{\epsilon_s}^s=\nu_s f_{\nu_s}$, where $ f_{\nu_s}$ is the observed radio flux\\
$f_{\epsilon_{EC}}^{EC}=\nu_{EC} f_{\nu_{EC}}$, where $ f_{\nu_{EC}}$ is the observed X-ray flux.\\

In the case of PKS 0637--752 ($z=0.651$ corresponding to $d_L=1.9 \, 10^{28}$ 
cm  or to $6.9$ kpc per arcsecond) the following values for the knot W7.8,  
taken by Schwartz et al. (2000) and by Chartas et al. (2000),
were used in \S \ref{section:power} (used also by DA04): 
 $\nu_s=4.8$ GHz, $ f_{\nu_s}=54 mJy$, $\alpha_r=0.8$,
$\nu_{EC}=3.8 \, 10^{17}$ Hz, $ f_{\nu_{EC}}=6.6\,10^{-9}$ Jy. The knot W7.8 
is not resolved in the optical and it is assumed that $r_b=1 $ Kpc, 
corresponding to a knot diameter of $\sim 0.3 ``$.

For 3C 273, ($z=0.158$ corresponding to $d_L=2.3 \, 10^{27}$ 
cm  or to $2.7$ kpc per arcsecond) the following values for  knot A,  
taken by Marshall et al. (2000)
were used in \S \ref{section:power}: 
 $\nu_s=1.65$ GHz, $ f_{\nu_s}=0.42 Jy$, $\alpha_r=0.76$,
$\nu_{EC}=2.4 \, 10^{17}$ Hz, $ f_{\nu_{EC}}=3.8 \,10^{-8}$ Jy. For knot A, 
$r_b=1 $ Kpc is assumed , corresponding to a knot diameter of $\sim 0.75 ``$.

\clearpage

\begin{figure}
\epsscale{0.8}
\plotone{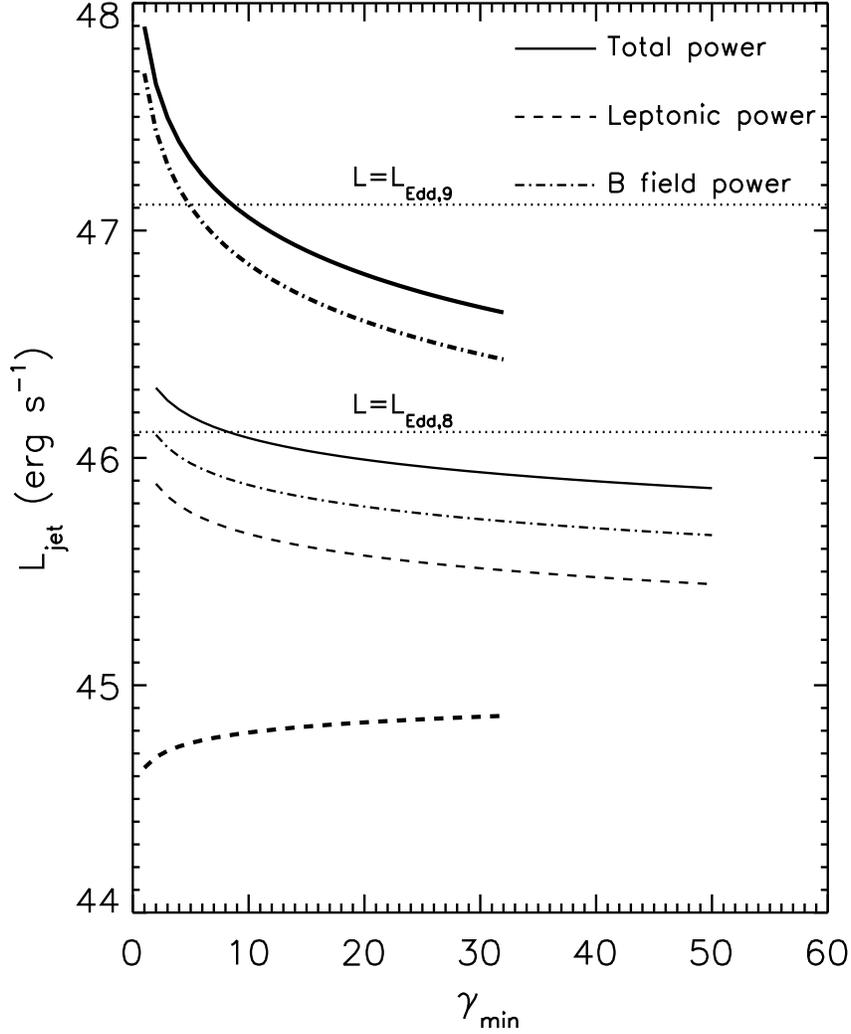}
\caption{The minimum power of a leptonic (hadronic) jet required by
 the knot  Wk7.8 of PKS 0637--752 is plotted with a solid thin (thick) 
line as a function of the permitted values of
the Lorentz factor $\gamma_{min}$ of the EED low energy cutoff.
The power carried by  radiating leptons in a minimum power configuration
is  shown with a broken thin (thick) line for a leptonic (hadronic) 
composition.  The thin (thick) dot-dash  lines show the Poynting flux in 
the case of a leptonic (hadronic) jet, also in minimum power conditions.
 The observational data needed in calculating
 $f_1$ in Equation (\ref{delta}) are those used by DA04 and are 
taken by Schwartz  et al. (2000) and Chartas et al. (2001). 
The two dotted lines show the Eddington luminosity that corresponds to
a $10^8\; M_{\odot}$ (lower dotted line) and $10^9\; M_{\odot}$. }
\label{power}
\end{figure}

\begin{thebibliography}{}

\bibitem[Aharonian (2002)]{aharonian02} Aharonian, F. A. 2002, \mnras, 332, 215

\bibitem[Begelman \& Sikora (1987)]{begelman1987}       Begelman, M. C. \& Sikora, M. 1987, \apj, 322, 650 

\bibitem[Blandford \& Znajek (1977)]{blandford77} Blandford, R. D. \&  Znajek, R. L. 1977, \mnras, 179, 433

\bibitem[Blandford \& Payne (1982)]{blandford82} Blandford, R. D. \&  Payne, D. G. \mnras, 1982, 199, 883

\bibitem[Celotti \&  Fabian (1993)]{celotti93}Celotti, A. \&  Fabian, A. C.  1993, \mnras,  264, 228

\bibitem[Celotti,  Ghisellini \&  Chiaberge (2001)]{celotti01}Celotti, A., Ghisellini, G., \&  Chiaberge M. 2001, \mnras,  321, L1

\bibitem[Chartas et al. (2001)]{chartas01} Chartas, G. et al. 2001, \apj, 542, 655

\bibitem[Cheung (2004)]{cheung04} Cheung, C. C. 2004, \apj, 600, L23 

\bibitem[Dermer \& Atoyan (2002)]{dermer02} Dermer, C. D. \& Atoyan, A. M.   2002, \apj, 568, L81

\bibitem[Dermer & Atoyan (2004a)]{dermer04a} Dermer, C D. \& Atoyan, A. 2004a,  \apj, 
611, L9 (DA04)

\bibitem[Dermer & Atoyan (2004b)]{dermer04b} Dermer, C D. \& Atoyan, A. 2004b,  \apj, 
613, 151 


\bibitem[Foley \& Davis (1985)]{foley85}
Foley, A. R. \& Davis, R. J. 1985, \mnras, 216, 679

\bibitem[Gallant (2002)]{gallant02} Gallant, Y. A.  2002, in
Relativistic Flows in Astrophysics, eds.  A.W. Guthmann, M. Georganopoulos, A. Marcowith, \& K. Manolakou,  Lecture Notes in Physics, vol. 589.

\bibitem[Georganopoulos \& Kazanas (2003)]{georganopoulos03} Georganopoulos, M. \& Kazanas, D. 2003, \apj, 589, L5

\bibitem[Georganopoulos \& Kazanas (2004)]{georganopoulos04} Georganopoulos, M. \& Kazanas, D. 2004, \apj, 604, L81

\bibitem[Ghisellini \& Celotti (2001)]{ghisellini01} Ghisellini, G., \& Celotti, A. 2001, \mnras, 327, 739

\bibitem[Gizani \& Leahy (2004)]{gizani04} Gizani, A. B. N. \& Leahy, J. P. 2004, \mnras, 350, 865

\bibitem[Hirotani (2005)]{hirotani04} Hirotani, K. 2005, \apj, 619, 73

\bibitem[Jester et al. (2002)]{jester02}Jester, S., R\"oser, H.-J., Meisenheimer, K., \&  Perley, R. 2002, \aap, 385, L27

\bibitem[Jorstad et al. (2002)]{jorstad02} Jorstad, S. G., Marscher, A. P., Mattox, J. R., Aller, M. F., Aller, H. D.,  Wehrle, A. E., \& Bloom, S. D. 2002, \apj, 556, 738

\bibitem[Jorstad \& Marscher (2004)]{jorstad04}Jorstad, S. G. \& Marscher, A. P. 2004, \apj, 614, 615

\bibitem[Kataoka \& Stawarz (2004)]{kataoka04} Kataoka, J, \& Stawarz, \L. 2004, \apj, in press, also in astro-ph/0411042

\bibitem[kirk et al. (2000)]{kirk00} Kirk, J. G., Guthmann, A. W., Gallant, Y. A., \& Achterberg, A. \apj,  542, 235 

\bibitem[K\"onigl (1989)] {konigl89} K\"onigl A. 1989, \apj, 342, 208 

\bibitem[Lister (2003)]{lister03} Lister, M. L. 2003, ApJ, 599, L105

\bibitem[Lovell et al. (2000)]{lovell00} Lovell, J. E. J. et al. 2000, in Astrophysical Phenomena Revealed by Space VLBI, ed. H. Hirabayashi, P. G. Edwards, \& D. W. Murphy (Sagamihara: ISAS), 215

\bibitem[Marshall et al. 2001]{marshall01} Marshall, H. L. et al. 2001, \apj, 549, L167

\bibitem[Marshall et al. 2005]{marshall05} Marshall, H. L. et al. 2005, \apjs, 156, 13

\bibitem[Martel et al. 2003]{martel03} Martel, A. R. et al. 2003, \aj, 125, 2964

\bibitem[McNamara et al. 2005]{mcnamara03} McNamara, B. R., Nulsen, P. E. J.,
Wise,  M. W., Rafferty,  D. A., Carilli, C., Sarazin, C. L., \& Blanton, E. L. 2005, \nat, 433, 45

\bibitem[Moderski et al. (2004)]{moderski04} Moderski, R., Sikora, M.,  
Madejski, G. M.,  \& Kamae, T. 2004, \apj, 611, 770

\bibitem[Omma \&  Binney (2004)]{omma04} Omma, H.  \& Binney, J. 2004, \mnras, 350, L13

	
\bibitem[Pearson et al. (1981)]{pearson81} Pearson, T. J. et al. 1981, \nat, 290, 365

\bibitem[Reynolds et al. (1996)]{reynolds91} Reynolds, C. S., Fabian, A. C., Celotti, A. \& Rees, M. J. 1996, \mnras, 283, 873

\bibitem[Rybicki \& Lightman (1979)]{rl79} Rybicki, G. R. \& Lightman, A. P. 1979, Radiative Processes in Astrophysics (Willey: New York)

\bibitem[Sambruna et al. 2001]{sambruna01} Sambruna, R. M., Urry, M. C., Tavecchio, F., Maraschi, L., Scarpa, R., Chartas, G., \& Muxlow, T.  2001, \apj, 549, L161

\bibitem[Sambruna et al. 2002]{sambruna02} Sambruna, R. M.,  Maraschi, L., Tavecchio, F., Urry, M. C., Cheung, C. C., Chartas, G.,  Scarpa, R., \& Gambill, J. K.      2002, \apj, 571, 206

\bibitem[Sambruna et al. 2004]{sambruna04} Sambruna,  R. M., Gambill, J. K.,  Maraschi, L., Tavecchio, F., Cerutti, R.,   Cheung, C. C., Urry, M. C., \& Chartas, G.    2004, \apj, 608, 698

\bibitem[Schwartz et al. 2000]{schwartz2000} Schwartz, D. E. et al. 2000, ApJ, 540, L69

\bibitem[Schwartz  (2002)]{schwartz02} Schwartz, D. E.  2002, \apj, 569, L23

\bibitem[Siemiginowska et al. (2002)]{siemiginowska02} Siemiginowska, A., Bechtold, J., Aldcroft, T. L., Elvis, M., Harris, D. E., \& Dobrzycki, A.  2002, \apj, 570, 543

\bibitem[Siemiginowska et al. (2003)] {siemiginowska03} Siemiginowska, A., Smith R. K., Aldcroft T. L., Schwartz, D. A., Paerels, F., \& Petric, A. O. 2003, \apj, 598, L15

\bibitem[Sikora et al. (1997)]{sikora97} Sikora, M., Madejski, G., Moderski, R. \& Poutanen, J. 1997, \apj, 484, 108

\bibitem[Sikora \& Madejski (2000)]{sikora00} Sikora, M. \&  Madejski, G. 2000, \apj, 534, 109
	
\bibitem[Spregel (2003)]{spregel03} Spergel, D. N. et al. 2003, \apjs, 148, 175



\bibitem[Stawarz et al. (2004)]{stawarz04} Stawarz, \L., Sikora, M., Ostrowski, M., \& Begelman, M. C. 2004, \apj, 608, 95 

\bibitem[Subrahmanyan, Saripalli, \& Hunstead (1996)]{sub96} Subrahmanyan, R., Saripalli, L., \& Hunstead, R. W. 1996, \mnras, 279, 257

\bibitem[Tavecchio et al. (2000)]{tavecchio00}   Tavecchio, F., Maraschi, L., Sambruna, R. \& Urry, C. M., 2000, \apj, 544, L23

\bibitem[Tavecchio, Ghisellini, \& Celotti (2003)]{tavecchio03} Tavecchio, F.,  Ghisellini, G., \& Celotti, A. 2003, \aap, 403, 83

	
\bibitem[Tavecchio et al. (2004)]{taveccio04}Tavecchio, F., Maraschi, L.,
 Sambruna, R. M., Urry, C. M., Cheung, C. C., Gambill, J. K., \& Scarpa, R.
2004, \apj, 614, 64

\bibitem[Wardle et al. (1998)] {wardle98}	 Wardle, J. F. C., Homan, D. C., Ojha, R., \& Roberts, D. H.  1998, \nat, 395, 457

\bibitem[Urry \& Padovani (1995)]{urry95}
Urry, C. M. \& Padovani, P. 1995,  \pasp, 107, 803

\bibitem[Yuan et al. (2003)]{yuan03} Yuan, W., Fabian, A. C., Celotti, A., \& Jonker, P. G. 2003, \mnras, 346, L7



\end{thebibliography}
\end{document}